# Weighted Impact Factor (WIF) for assessing the quality of scientific journals

**Rasim Alguliyev**
Institute of Information Technology, Azerbaijan National Academy of Sciences, Azerbaijan

**Ramiz Aliguliyev**
Institute of Information Technology, Azerbaijan National Academy of Sciences, Azerbaijan

**Nigar Ismayilova**
Institute of Information Technology, Azerbaijan National Academy of Sciences, Azerbaijan

## Abstract

Nowadays impact factor is the significant indicator for journal evaluation. In impact factor's calculation is used number of all citations to journal, regardless of the prestige of cited journals, however, scientific units (paper, researcher, journal or scientific organization) cited by journals with high impact factor or researchers with high Hirsch index are more important than objects cited by journals without impact factor or unknown researcher. In this paper was offered weighted impact factor for getting more accurate rankings for journals, which consider not only quantity of citations, but also quality of citing journals. Correlation coefficients among different indicators for journal evaluation: impact factors by Thomson Scientific, weighted impact factors offered by different researchers, average and medians of all citing journals' impact factors and 5-year impact factors were analysed.

## 1. Introduction

Today, a lot of scientific papers in different subject areas, fields are published. So determining the influence of each paper or journal where these papers are published is very important. Nowadays the most used indicator for journal influence measuring is impact factor proposed by Garfield [1]. The generally accepted as an indicator of journal prestige the Thomson Reuters (earlier Institute of Scientific Information - ISI) calculates the impact factor of $j^{th}$ journal in year $t$ as ratio of number of citations in current year to articles published in this journal during previous two years to number of these articles:

$$IF_j^t = \frac{\sum_{i=1}^{n_j^t} c_{ij}^t}{a_j^{t-1} + a_j^{t-2}},$$

$$(1)$$

where $n_j^t$ – is the number of journals in the set of all journals indexed in year $t$ cited the articles of journal, $c_{ij}^t$ – is the number of citations to journal $j$ from journal $i$ in year $t$ and $a_j^t$ – is the total number of articles published in year $t$. Although IF is an important quotient for evaluation of scientific journals used by librarians, researchers, science policymakers, there are many critiques against the IF efficiency [2-3]. Reedjik coordinates it by wrong, biased and even manipulated citations, as a result of citation habits for authors in different fields, selectivity in citations by authors, errors made by authors in citation lists at the end of documents and by ISI in entering publications and citations in databases, in classifying citations and accrediting them to journals and authors [4]. During the last decades were offered

**Corresponding author:**
Ramiz Aliguliyev, Institute of Information Technology, Azerbaijan National Academy of Sciences, 9 B. Vahabzade Street, AZ1141, Baku, Azerbaijan
E-mail: r.aliguliyev@gmail.com

different modifications of ISI IF. For using with complex citation databases as Scopus was proposed the SCImago Journal Rank on citation weighting schemes and eigenvector centrality was proposed [5]. Perez-Hornero et al. proposed a Bayesian approach to the problem with taking into account journals' recent trajectory besides the current prestige of given journal [6]. Although during the last decades were proposed new indicators for journals evaluation, was concluded that, they (the Eigenvector, the SCImago journal rank, an Article Influence Score and journal h-index) correlate very much with the IF proposed by ISI and among each other [7]. One of the main disadvantages of IF is the equality of citations, regardless of the importance of the citing journals [8-10]. Unfortunately, in calculation of IF prestige, reputation of citing journals are not considered, nevertheless 10 citations from journal with high IF must be preferred to by one citation from 10 journals with low IF or without IF. For solution of this problem in this paper was proposed impact factor calculated by taking into 5-year impact factor of each citing journal.

## 2. Related Work

In general, journals in specialized and applied disciplines get more citations than journals in fundamental subject areas. The numbers of researchers and publishing journals in each area greatly affect the impact factor [11]. In this reason some researchers proposed methods for normalization of impact factors of journals for different scientific areas. Owlia et al. introduced normalized impact factor for evaluating the quality of journals and research works in different disciplines [12]. The normalized IF (NIF) index was established based on multiplication of journal IF by constant factor. NIF is the ratio of the aggregate impact factors of journals to aggregate impact factor of each discipline in given scientific field:

$$NIF^t = \frac{C^t J^t}{A^t \check{C}^t},$$                                                                                                                (2)

where $C^t$ - is the total number of citations to the entire given scientific field in current year, $A^t$ - is the total number of articles published by journals from entire given scientific field in current year, $\check{C}^t$ - is the total number of citations to the journals of the particular discipline and $J^t$ - is the total number of journals of the particular discipline published articles. Other method for normalization of journal impact factors was produced by Iftikhar, Masood and Song [13]. Modified impact factors were calculated at disciplines, branches and specialities level and designated as Red, Yellow and Green MIF respectively. For modification the highest ISI IF of journal of that discipline, branch or speciality is weighted as 100 and other groups members are normalized accordingly by considering it as reference point. Also was proposed rank-normalized impact factor [14], Source Normalized Impact per Paper (SNIP) by Scopus [15].

Weighting citations for more effective impact factor and more accurate ranking of journals have also been suggested by some researchers. In 1976 was developed self-consistent methodology for determining citation influence weights for scientific journals, fields and subfields. Also was proposed modification of IF using Weighted PageRank [16]. Authors suggested a simple combination of both the IF and the weighted PageRank and found that the resulting journal rankings correspond well to a general understanding of journal status [17]. Zyezkowski formulated weighted impact factor of journals using weighted citations, weighted impact factor of papers, weighted Hirsch index and weighted efficiency index [18]. In order to not ignoring the impact or prestige of the cited journals were proposed different modifications of impact factor. Buela-Casal had offered two indices: a mean impact factor of the citing journals (MIFCJ) and a weighted impact factor (WIF). MIFCJ is quotient of multiplication of impact factors of citing journals by the number of cited articles to the total number citing articles [19]:

$$MIFJC_j^t = \frac{\sum_{i=1}^{n_j^f} IF_j^{t-1} c_{ij}^t}{a_j^{t-1} + a_j^{t-2}},$$                                                                                       (3)

Weighted impact factor of a journal is the average of MIFCJ and IF from ISI Journal Citation Ranking (JCR) reports:

$$WIF_j^t = \frac{MIFJC_j^t + IF_j^{t-1}}{2}.$$                                                                                                        (4)

Habibzadeh and Yadollahie suggested weighted impact factor with relative weights of journals by their IF in the previous year normalized using logistic function [20]:

$$WIF_j^t = \frac{\sum_{i=1}^{n_j^t} w_{ij}^t c_{ij}^t}{a_j^{t-1} + a_j^{t-2}}, \tag{5}$$

where $w_{ij}^t$ designates the weight of relative to $J_i$ weighted impact factor of journal $J_k$ in year $t$:

$$w_{ij}^t = 10 \frac{1 - 0.828 e^{-q_{ij}^t}}{1 + 16.183 e^{-q_{ij}^t}}, \tag{6}$$

where $q_{ij}^t$ - is a quotient the numerator of which is the IF of citing journal in the previous year and the denominator is IF of cited journal in the previous year:

$$q_{ij}^t = \frac{IF_i^{t-1}}{IF_j^{t-1}}, \tag{7}$$

Although proposed by researchers weighted IF takes into account not only the number of times that the journal has been cited, but also the prestige of the journals by which it has been cited, were determined some serious problems of proposed WIF (weighted impact factor). Because of ranking of journals based in this WIF can be misleading, was indicated how the problems with the WIF proposed by Habibzadeh and Yadollahie (H&Y) can be solved [21].

In our opinion, the main lack of WIF proposed by H&Y is dependence on IF of cited journal in previous year.

**Example**: let us calculate weighted impact factors of 3 journals for current year with the same number of articles in previous two years, we assume that number of citations and impact factors of citing journal to these articles in current year are the same for each journal (table 1):

      Number of articles: $P_i = 20$;

      Impact factors of journals in previous year: $IF_1 = 2$; $IF_2 = 4$; $IF_3 = 6$;

**Table 1.** Impact factors and numbers of citations of citing journals for journal in example.

| N | IF of citing journal | Number of citations |
|---|---|---|
| 1 | 2 | 2 |
| 2 | 4 | 2 |
| 3 | 8 | 2 |
| 4 | 16 | 2 |

      Using data from table 1 were calculated weighted impact factors for given journals proposed by H&Y:

      $WIF_1 = 0.994324$; $WIF_2 = 0.759668$; $WIF_3 = 0.443629$

As result we can see that, calculated weighted impact factors for 3 journals with the same number of citations from the same sources are different depending on impact factors of journals in previous year. Cited journals with high impact factor in previous year gets lower weighted impact factors. This disadvantage of proposed IF makes ranking of journals by this method undesirable.

## 3. Proposed weighted impact factor

Assume that $S^t$ − is the set of all journals indexed in year $t$, $n^t$ − is the number of indexed journals, $n_j^t$ − is the number of journals in $S^t$ cited the articles of journal j, $a_j^t$ − is the total number of articles published in year $t$, $c_{ij}^t$ − is the number of citations to journal $j$ from journal $i$ in year $t$, $FIF_j^t$ − is 5-year impact factor of journal $j$ in year $t$. Proposed weighted IF of journal $j$ in year $t$ is

$$WIF_j^t = \frac{\sum_{i=1}^{n_j^t} \left( FIF_j^{t-1} + 1 \right) c_{ij}^t}{a_j^{t-1} + a_j^{t-2}}, \tag{7}$$

To analyse the efficiency of proposed impact factor, we tried to calculate new impact factors in 2013 for selected twenty journals in computer science sphere from JCR for 2013 (Table 2).

**Table 2.** Some indicators of analysed journals from Scopus Database.

| N | Journal | Number of articles published in 2011 and 2012 | Number of citations in 2013 to articles published in 2011 and 2012 | Impact factor in 2012 |
|---|---|---|---|---|
| 1 | Neural Computation | 226 | 383 | 1.76 |
| 2 | Swarm Intelligence | 26 | 48 | 0.64 |
| 3 | Neural Processing Letters | 76 | 94 | 1.24 |
| 4 | Artificial Life | 48 | 93 | 1.585 |
| 5 | Cognitive Computation | 88 | 97 | 0.867 |
| 6 | Computer Speech And Language | 67 | 121 | 1.463 |
| 7 | Fuzzy Optimization and Decision Making | 45 | 45 | 1.488 |
| 8 | Genetic Programming and Evolvable Machines | 42 | 45 | 1.333 |
| 9 | International Journal of Appled Mathematics and Computer Science | 136 | 189 | 1.008 |
| 10 | Journal of Ambient Intelligence and Smart Environments | 74 | 80 | 1.298 |
| 11 | ACM Transactions on Applied Perception | 40 | 42 | 1 |
| 12 | ACM Transactions on Knowledge Discovery from Data | 37 | 42 | 1.676 |
| 13 | Acm Transactions on Information Systems | 42 | 55 | 1.07 |
| 14 | ACM Transactions on the Web | 39 | 62 | 1.405 |
| 15 | ACM Transactions on Sensor Networks | 54 | 79 | 1.444 |
| 16 | Acm Transactions on Software Engineering And Methodology | 37 | 54 | 1.548 |
| 17 | IEEE Transactions on Computational Intelligence and AI in Games | 50 | 58 | 1.694 |
| 18 | IEEE Transactions on Dependable and Secure Computing | 143 | 163 | 1.059 |
| 19 | IEEE Transactions on Autonomous Mental Development | 54 | 73 | 2.17 |
| 20 | World Wide Web | 58 | 94 | 1.196 |

For these journals number of articles published in 2011 and 2012, total number of citations in 2013, number of sources citing journal were determined using data about articles and their citations from Scopus database. Also the numbers of citations from each journal with impact factors were determined for given journals (Table 3).

**Table 3.** 5-year impact factors of citing journals and number of citations from them.

### J1 – Neural Computation

| 1 | 2 | 3 | 4 | 5 | 6 | 7 | 8 | 9 |
|---|---|---|---|---|---|---|---|---|
| 35.89 (1) | 34.37 (1) | 31.03 (3) | 23.17 (1) | 16.41 (1) | 16.40 (4) | 13.58 (1) | 14.57 (1) | 14.47 (1) |
| 13.45 (1) | 11.34 (3) | 10.58 (9) | 10.45 (1) | 9.924 (2) | 7.87 (10) | 7.463 (2) | 7.063 (4) | 6.895 (3) |
| 6.144 (5) | 6 (1) | 5.94 (14) | 5.484 (1) | 4.885 (1) | 4.544 (1) | 4.479 (4) | 4.422 (1) | 4.284 (5) |
| 4.25 (1) | 4.24 (11) | 4.049 (7) | 4.017 (1) | 3.879 (2) | 3.844 (1) | 3.71(1) | 3.707 (1) | 3.676 (1) |
| 3.668 (2) | 3.646 (1) | 3.632 (9) | 3.612 (2) | 3.607 (8) | 3.568 (1) | 3.291 (1) | 3.219 (1) | 3.146 (1) |
| 3.108 (1) | 3.069 (1) | 3.068 (3) | 3.05 (1) | 2.998 (3) | 2.927 (2) | 2.895 (1) | 2.892 (4) | 2.743 (1) |
| 2.733 (1) | 2.653 (2) | 2.61 (20) | 2.567 (1) | 2.526 (1) | 2.525 (1) | 2.5 (10) | 2.496 (1) | 2.484 (2) |
| 2.38 (31) | 2.339 (1) | 2.307 (3) | 2.287 (3) | 2.27 (1) | 2.158 (1) | 2.143 (2) | 2 (1) | 1.947 (1) |
| 1.938 (3) | 1.922 (3) | 1.871 (1) | 1.811 (8) | 1.767 (1) | 1.745 (1) | 1.732 (1) | 1.724 (1) | 1.643 (1) |
| 1.6 (1) | 1.596 (1) | 1.595 (1) | 1.572 (1) | 1.55 (1) | 1.529 (2) | 1.402 (2) | 1.386 (2) | 1.338 (1) |
| 91 | 92 | 93 | 94 | 95 | 96 | 97 | 98 | 99 |
| 1.336 (1) | 1.319 (1) | 1.314 (1) | 1.305 (1) | 1.231(1) | 1.216 (1) | 1.192 (1) | 1.183 (1) | 1.182 (1) |
| 100 | 101 | 102 | 103 | 104 | 105 | 106 | 107 | 108 |
| 1.14 (3) | 1.074 (1) | 1.032 (1) | 0.97 (1) | 0.84 (1) | 0.816 (1) | 0.697 (1) | 0.672 (2) | 0.6 (1) |
| 109 | 110 | 111 | | | | | | |
| 0.548 (1) | 0.297 (1) | 0.187 (2) | | | | | | |

## J2 – Swarm Intelligence

| | | | | | | | | |
|---|---|---|---|---|---|---|---|---|
| 1 | 2 | 3 | 4 | 5 | 6 | 7 | 8 | 9 |
| 6.226 (1) | 5.165 (2) | 3.448 (1) | 3.405 (1) | 3.097 (2) | 2.747 (1) | 1.957 (1) | 1.615 (1) | 1.545 (2) |
| 10 | 11 | | | | | | | |
| 1.364 (1) | 0.953 (2) | | | | | | | |

## J3 – Neural Processing Letters

| | | | | | | | | |
|---|---|---|---|---|---|---|---|---|
| 1 | 2 | 3 | 4 | 5 | 6 | 7 | 8 | 9 |
| 7.854 (1) | 4.268 (1) | 3.676 (1) | 3.632 (2) | 3.513 (1) | 3.219 (1) | 2.501 (2) | 2.457 (3) | 2.384 (2) |
| 10 | 11 | 12 | 13 | 14 | 15 | 16 | 17 | 18 |
| 2.339 (1) | 2.143 (1) | 1.831 (1) | 1.811 (8) | 1.71 (1) | 1.529 (1) | 1.454 (2) | 1.42 (1) | 1.36 (1) |
| 19 | 20 | 21 | 22 | 23 | 24 | 25 | 26 | 27 |
| 1.329 (1) | 1.23 (10) | 1.216 (1) | 1.183 (2) | 1.074 (6) | 1.04 (3) | 0.898 (1) | 0.866 (1) | 0.84 (2) |
| 28 | 29 | 30 | 31 | 32 | 33 | 34 | 35 | 36 |
| 0.774 (5) | 0.755 (1) | 0.753 (1) | 0.716 (1) | 0.682 (5) | 0.64 (1) | 0.622 (1) | 0.561 (1) | 0.497 (2) |

## J4 – Artificial Life

| | | | | | | | | |
|---|---|---|---|---|---|---|---|---|
| 1 | 2 | 3 | 4 | 5 | 6 | 7 | 8 | 9 |
| 17.72 (1) | 13.56 (1) | 7.51 (1) | 7.435 (1) | 6.69 (1) | 6.226 (1) | 5.165 (2) | 4.728 (1) | 4.446 (1) |
| 10 | 11 | 12 | 13 | 14 | 15 | 16 | 17 | 18 |
| 4.406 (1) | 4.244 (7) | 2.496 (1) | 2.333 (1) | 2.307 (1) | 2 (1) | 1.945 (1) | 1.777 (1) | 1.545 (2) |
| 19 | 20 | 21 | 22 | 23 | 24 | 25 | 26 | |
| 1.454 (1) | 1.364 (1) | 1.336 (1) | 0.953 (2) | 0.816 (1) | 0.617 (1) | 0.48 (1) | 0.417 (1) | |

## J5 –Cognitive Computation

| | | | | | | | | |
|---|---|---|---|---|---|---|---|---|
| 1 | 2 | 3 | 4 | 5 | 6 | 7 | 8 | 9 |
| 9.924 (1) | 7.869 (1) | 7.298 (1) | 4.479 (1) | 4.372 (1) | 4.244 (3) | 4.017 (1) | 3.674 (1) | 3.598 (1) |
| 10 | 11 | 12 | 13 | 14 | 15 | 16 | 17 | 18 |
| 3.262 (1) | 2.998 (2) | 2.847 (1) | 2.538 (1) | 2.525 (1) | 2.501 (3) | 2.445 (1) | 2.339 (1) | 2.194 (1) |
| 19 | 20 | 21 | 22 | 23 | 24 | 25 | 26 | 27 |
| 2.158 (1) | 1.938 (2) | 1.936 (1) | 1.811 (2) | 1.745 (1) | 1.596 (1) | 1.529 (2) | 1.52 (1) | 1.423 (1) |
| 28 | 29 | 30 | 31 | 32 | 33 | | | |
| 1.157 (2) | 1.14 (14) | 0.846 (1) | 0.735 (1) | 0.592 (1) | 0.326 (1) | | | |

## J6 – Computer Speech And Language

| | | | | | | | | |
|---|---|---|---|---|---|---|---|---|
| 1 | 2 | 3 | 4 | 5 | 6 | 7 | 8 | 9 |
| 7.694 (2) | 2.643 (1) | 2.395 (2) | 2.339 (2) | 1.952 (1) | 1.936 (1) | 1.915 (2) | 1.708 (1) | 1.52 (8) |
| 10 | 11 | 12 | 13 | 14 | 15 | 16 | 17 | 18 |
| 1.423 (5) | 1.41 (1) | 1.146 (1) | 1.137 (4) | 1.074 (2) | 0.977 (1) | 0.959 (1) | 0.932 (1) | 0.767 (1) |
| 19 | 20 | 21 | 22 | 23 | | | | |
| 0.664 (1) | 0.617 (1) | 0.505 (1) | 0.466 (1) | 0.305 (1) | | | | |

## J7 – Fuzzy Optimization and Decision Making

| | | | | | | | | |
|---|---|---|---|---|---|---|---|---|
| 1 | 2 | 3 | 4 | 5 | 6 | 7 | 8 | 9 |
| 3.676 (3) | 2.218 (2) | 2.167 (1) | 2.165 (1) | 1.81 (15) | 1.721 (1) | 1.674 (2) | 1.579 (2) | 1.386 (1) |
| 10 | 11 | 12 | 13 | 14 | 15 | | | |
| 1.364 (4) | 1.183 (2) | 0.846 (2) | 0.746 (1) | 0.612 (1) | 0.27 (16) | | | |

## J8 – Genetic Programming and Evolvable Machines

| | | | | | | | | |
|---|---|---|---|---|---|---|---|---|
| 1 | 2 | 3 | 4 | 5 | 6 | 7 | 8 | 9 |
| 3.676 (1) | 3.027 (1) | 2.526 (1) | 2.501 (1) | 1.938 (1) | 1.811 (1) | 1.795 (1) | 1.726 (2) | 1.625 (1) |
| 10 | 11 | 12 | 13 | | | | | |
| 1.39 (1) | 1.349 (1) | 1.282 (8) | 1.231 (1) | | | | | |

## J9 – International Journal of Appled Mathematics and Computer Science

| | | | | | | | | |
|---|---|---|---|---|---|---|---|---|
| 1 | 2 | 3 | 4 | 5 | 6 | 7 | 8 | 9 |
| 4.244 (1) | 3.676 (1) | 3.601(1) | 3.212 (1) | 2.64 (2) | 2.62 (1) | 2.457 (1) | 2.382 (1) | 2.255 (1) |
| 10 | 11 | 12 | 13 | 14 | 15 | 16 | 17 | 18 |
| 2.151 (1) | 2.04 (4) | 1.758 (1) | 1.674 (1) | 1.651 (1) | 1.625 (1) | 1.529 (1) | 1.504 (1) | 1.454 (1) |
| 19 | 20 | 21 | 22 | 23 | 24 | 25 | 26 | 27 |
| 1.368 (2) | 1.364 (2) | 1.359 (1) | 1.289 (2) | 1.216 (3) | 1.201 (2) | 1.183 (1) | 1.182 (1) | 1.158 (1) |

| 28 | 29 | 30 | 31 | 32 | 33 | 34 | 35 | 36 |
|---|---|---|---|---|---|---|---|---|
| 1.15 (52) | 1.024 (1) | 0.932 (1) | 0.898 (2) | 0.829 (1) | 0.8 (4) | 0.671 (1) | 0.61 (1) | 0.594 (3) |
| 37 | 38 | 39 | 40 | 41 | 42 | 43 | | |
| 0.548 (1) | 0.483 (1) | 0.436 (1) | 0.41 (1) | 0.395 (1) | 0.37 (1) | 0.269 (1) | | |

**J10 – Journal of Ambient Intelligence and Smart Environments**

| 1 | 2 | 3 | 4 | 5 | 6 | 7 | 8 | 9 |
|---|---|---|---|---|---|---|---|---|
| 3.382 (1) | 2.7 (1) | 2.632 (1) | 2.525 (1) | 2.339 (1) | 2.003 (1) | 1.947 (1) | 1.811 (2) | 1.64 (16) |
| 10 | 11 | | | | | | | |
| 1.529 (1) | 1.169 (1) | | | | | | | |

**J11 – ACM Transactions on Applied Perception**

| 1 | 2 | 3 | 4 | 5 | 6 | 7 | 8 | 9 |
|---|---|---|---|---|---|---|---|---|
| 6.144 (1) | 4.283 (2) | 4.017 (1) | 2.62 (1) | 2.61 (1) | 2.566 (1) | 2.445 (2) | 2.395 (1) | 2.292 (1) |
| 10 | 11 | 12 | 13 | 14 | 15 | 16 | 17 | 18 |
| 2.007 (1) | 2 (1) | 1.905 (1) | 1.36 (2) | 1.269 (5) | 1.216 (1) | 1.112 (1) | 0.675 (1) | 0.5 (1) |
| 19 | | | | | | | | |
| 0.453 (1) | | | | | | | | |

**J12 – ACM Transactions on Knowledge Discovery from Data**

| 1 | 2 | 3 | 4 | 5 | 6 | 7 | 8 | 9 |
|---|---|---|---|---|---|---|---|---|
| 4.395 (1) | 4.244 (1) | 4.017 (1) | 3.959 (1) | 3.676 (3) | 3.371 (1) | 3.263 (1) | 3.068 (1) | 2.426 (1) |
| 10 | 11 | 12 | 13 | 14 | 15 | | | |
| 2.339 (3) | 1.838 (1) | 1.811 (1) | 1.359 (2) | 0.739 (1) | 0.707 (1) | | | |

**J13 – ACM Transactions on Information Systems**

| 1 | 2 | 3 | 4 | 5 | 6 | 7 | 8 | 9 |
|---|---|---|---|---|---|---|---|---|
| 3.676 (1) | 3.371 (3) | 3.037 (2) | 2.566 (1) | 2.446 (1) | 2.339 (2) | 1.745 (1) | 1.716 (3) | 1.586 (1) |
| 10 | 11 | 12 | | | | | | |
| 1.318 (1) | 1.109 (2) | 0.466 (1) | | | | | | |

**J14 – ACM Transactions on the Web**

| 1 | 2 | 3 | 4 | 5 | 6 | 7 | 8 | 9 |
|---|---|---|---|---|---|---|---|---|
| 7.854 (2) | 4.395 (1) | 3.676 (2) | 3.371 (1) | 3.037 (2) | 2.927 (1) | 2.446 (5) | 2.424 (1) | 2.339 (3) |
| 10 | 11 | 12 | 13 | 14 | 15 | 16 | 17 | 18 |
| 2.158 (3) | 2.033 (1) | 1.469 (1) | 1.452 (1) | 1.388 (2) | 1.384 (1) | 1.322 (2) | 1.109 (1) | 0.943 (2) |
| 19 | 20 | 21 | | | | | | |
| 0.785 (1) | 0.765 (1) | 0.605 (1) | | | | | | |

**J15 – ACM Transactions on Sensor Networks**

| 1 | 2 | 3 | 4 | 5 | 6 | 7 | 8 | 9 |
|---|---|---|---|---|---|---|---|---|
| 6.348 (2) | 6.146 (1) | 3.587 (2) | 3.371 (1) | 2.747 (1) | 2.485 (1) | 2.395 (1) | 2.203 (1) | 1.957 (1) |
| 10 | 11 | 12 | 13 | 14 | 15 | 16 | 17 | 18 |
| 1.859 (1) | 1.758 (2) | 1.227 (1) | 1.183 (4) | 1.169 (1) | 1.092 (1) | 1.002 (1) | 0.765 (1) | 0.605 (1) |
| 19 | | | | | | | | |
| 0.43 (5) | | | | | | | | |

**J16 – ACM Transactions on Software Engineering And Methodology**

| 1 | 2 | 3 | 4 | 5 | 6 | 7 | 8 | 9 |
|---|---|---|---|---|---|---|---|---|
| 3.612 (1) | 3.371 (1) | 2.063 (3) | 2.031 (2) | 1.756 (1) | 1.692 (2) | 1.322 (1) | 1.167 (4) | 1.867 (1) |
| 10 | 11 | 12 | 13 | 14 | 15 | 16 | 17 | 18 |
| 0.819 (1) | 0.785 (1) | 0.727 (1) | 0.721 (1) | 0.682 (2) | 0.43 (1) | 0.336 (2) | 0.269 (1) | 0.268 (1) |

**J17 – IEEE Transactions on Computational Intelligence and AI in Games**

| 1 | 2 | 3 | 4 | 5 | 6 | 7 | 8 | 9 |
|---|---|---|---|---|---|---|---|---|
| 3.212 (1) | 3.071 (1) | 2.339 (2) | 1.936 (1) | 1.63 (11) | 1.282 (1) | 0.954 (1) | 0.832 (1) | 0.723 (3) |

**J18 – IEEE Transactions on Dependable and Secure Computing**

| 1 | 2 | 3 | 4 | 5 | 6 | 7 | 8 | 9 |
|---|---|---|---|---|---|---|---|---|
| 6.895 (1) | 3.676 (4) | 3.371 (2) | 3.191 (1) | 3.071 (3) | 2.744 (1) | 2.426 (1) | 2.339 (2) | 2.259 (4) |
| 10 | 11 | 12 | 13 | 14 | 15 | 16 | 17 | 18 |
| 2.067 (1) | 2.021 (1) | 1.894 (1) | 1.726 (10) | 1.576 (2) | 1.42 (1) | 1.39 (1) | 1.341 (1) | 1.322 (2) |
| 19 | 20 | 21 | 22 | 23 | 24 | 25 | 26 | 27 |
| 1.317 (1) | 1.291 (1) | 1.234 (1) | 1.227 (1) | 1.106 (1) | 1.093 (3) | 1.019 (1) | 0.954 (2) | 0.945 (1) |
| 28 | 29 | 30 | 31 | 32 | 33 | 34 | 35 | 36 |
| 0.874 (1) | 0.867 (1) | 0.819 (1) | 0.793 (1) | 0.712 (1) | 0.701 (1) | 0.697 (3) | 0.625 (2) | 0.606 (1) |
| 37 | 38 | 39 | 40 | 41 | 42 | 43 | | |
| 0.605 (1) | 0.497 (2) | 0.463 (2) | 0.43 (3) | 0.395 (2) | 0.297 (1) | 0.187 (1) | | |

## J19 – IEEE Transactions on Autonomous Mental Development

| 1 | 2 | 3 | 4 | 5 | 6 | 7 | 8 | 9 |
|---|---|---|---|---|---|---|---|---|
| 4.244 **(4)** | 3.587 **(1)** | 2.501 **(4)** | 2.26 **(17)** | 2.202 **(1)** | 1.947 **(1)** | 1.859 **(1)** | 1.615 **(3)** | 1.545**(2)** |
| 10 | 11 | 12 | 13 | 14 | 15 | 16 | 17 | |
| 1.529 **(1)** | 1.423 **(1)** | 1.137 **(1)** | 1.043 **(1)** | 0.898 **(1)** | 0.817 **(1)** | 0.675 **(1)** | 0.567 **(1)** | |

## J20 – World Wide Web

| 1 | 2 | 3 | 4 | 5 | 6 | 7 | 8 | 9 |
|---|---|---|---|---|---|---|---|---|
| 3.676 **(1)** | 3.219 **(2)** | 3.191 **(1)** | 2.446 **(1)** | 2.426 **(1)** | 2.403 **(1)** | 2.339 **(2)** | 2.031 **(1)** | 1.955 **(1)** |
| 10 | 11 | 12 | 13 | 14 | 15 | 16 | 17 | 18 |
| 1.838 **(1)** | 1.586 **(1)** | 1.45 **(23)** | 1.251 **(1)** | 1.169 **(1)** | 0.94 **(1)** | 0.332 **(1)** | 0.302 **(1)** | 0.174 **(1)** |

here (*) denotes the number of citations

Weighted impact factors of given journals were compared with JCR IF for 2013, WIF proposed by H&Y, Buela-Casal and impact factors weighted by average and median of citing journals' impact factors and 5-year impact factors (Table 4).

**Table 4.** JCR IF, WIF by H&Y, by Buela-Casal, proposed WIF, average and median WIF, weighted by average and median of 5-year impact factors for analysed journals.

| N | Journal | JCR IF | WIF by H&Y | WIF by Buela-Casal | Proposed WIF | Average WIF | Median WIF | Average 5WIF | Median 5WIF |
|---|---|---|---|---|---|---|---|---|---|
| 1 | Neural Computation | 1.694 [4] | 4.200 [2] | 3.603 [1] | 7.544 [1] | 6.533 [1] | 4.200[2] | 7.725 [1] | 4.9 [3] |
| 2 | Swarm Intelligence | 1.833 [2] | 4.457 [1] | 1.577 [9] | 3.472 [4] | 3.238 [3] | 2.055 [8] | 5.204 [3] | 5.071 [2] |
| 3 | Neural Processing Letters | 1.237 [12] | 1.631 [12] | 1.455 [13] | 2.763 [8] | 1.765 [14] | 1.534 [15] | 2.02 [13] | 1.523 [17] |
| 4 | Artificial Life | 1.93 [1] | 3.028 [3] | 2.688 [2] | 4.762 [2] | 6.474 [2] | 4.518 [1] | 7.724 [2] | 8.223 [1] |
| 5 | Cognitive Computation | 1.1 [16] | 2.649 [4] | 1.317 [17] | 2.594 [11] | 2.253 [9] | 1.955 [10] | 2.631 [11] | 1.996 [12] |
| 6 | Computer Speech And Language | 1.812 [3] | 1.915 [8] | 1.778 [7] | 2.875 [6] | 2.137 [11] | 2.317 [4] | 3.091 [6] | 2.57 [8] |
| 7 | Fuzzy Optimization and Decision Making | 1 [20] | 0.855 [20] | 1.330 [16] | 2.158 [15] | 1.172 [20] | 1.124 [20] | 1.33 [20] | 1.48 [20] |
| 8 | Genetic Programming and Evolvable Machines | 1.065 [18] | 1.193 [18] | 1.363 [15] | 1.919 [18] | 1.729 [15] | 1.453 [16] | 1.88 [17] | 1.508 [18] |
| 9 | International Journal of Applied Mathematics and Computer Science | 1.39 [9] | 1.698 [11] | 1.266 [19] | 2.443 [12] | 1.643 [18] | 1.401 [18] | 1.815 [18] | 1.593 [16] |
| 10 | Journal of Ambient Intelligence and Smart Environments | 1.082 [17] | 1.198 [17] | 1.294 [18] | 1.758 [20] | 1.658 [17] | 1.403 [17] | 2.006 [14] | 1.773 [14] |
| 11 | ACM Transactions on Applied Perception | 1.051 [19] | 2.021 [7] | 1.264 [20] | 2.426 [13] | 1.821 [12] | 1.700 [12] | 2.223 [12] | 2.05 [11] |

| 12 | ACM Transactions on Knowledge Discovery from Data | 1.147 [14] | 1.573 [13] | 1.806 [6] | 2.61 [9] | 2.480 [6] | 2.105 [7] | 3.099 [5] | 3.118 [5] |
| 13 | ACM Transactions On Information Systems | 1.3 [11] | 2.059 [6] | 1.389 [14] | 2.3 [14] | 2.406 [7] | 2.132 [6] | 2.897 [9] | 3.063 [6] |
| 14 | ACM Transactions on the Web | 1.595 [6] | 2.076 [5] | 1.829 [5] | 3.828 [3] | 2.670 [4] | 2.234 [5] | 3.855 [4] | 3.718 [4] |
| 15 | ACM Transactions on Sensor Networks | 1.463 [8] | 1.697 [10] | 1.615 [8] | 2.607 [10] | 2.285 [8] | 1.612 [14] | 3.013 [7] | 2.184 [10] |
| 16 | ACM Transactions On Software Engineering And Methodology | 1.472 [7] | 1.369 [16] | 1.487 [12] | 2.098 [16] | 1.674 [16] | 1.660 [13] | 1.912 [15] | 1.707 [15] |
| 17 | IEEE Transactions on Computational Intelligence and AI in Games | 1.167 [13] | 1.139 [19] | 1.549 [10] | 1.88 [19] | 1.804 [13] | 1.965 [9] | 1.899 [16] | 1.885 [13] |
| 18 | IEEE Transactions on Dependable and Secure Computing | 1.137 [15] | 1.533 [14] | 1.195 [4] | 1.985 [17] | 1.534 [19] | 1.230 [19] | 1.812 [19] | 1.507 [19] |
| 19 | IEEE Transactions on Autonomous Mental Development | 1.348 [10] | 1.450 [15] | 2.209 [3] | 3.222 [5] | 2.531 [5] | 2.864 [3] | 2.929 [8] | 3.053 [7] |
| 20 | World Wide Web | 1.623 [5] | 1.862 [9] | 1.547 [11] | 2.832 [7] | 2.200 [10] | 1.938 [11] | 2.71 [10] | 2.353 [9] |

[*] denotes the rank of journals.

## 4. Discussion

For describing the efficiency of proposed method Pearson correlation coefficients were calculated between sets of indictors (JCR IF, WIF by H & Y, WIF by G. Buela-Casal, proposed WIF, average and median WIF, average and median 5WIF) (Table 5). Results are illustrated graphically in Figures 1 to 15. Analyzing correlation coefficients among different indicators for journal evaluation, we can notice that, proposed in this paper WIF have high correlations with all analyzed indicators, excluding WIF proposed by G. Buela-Casal, which has low correlations with all indicators in this list. The WIF proposed by H&Y and WIF proposed by G. Buela-Casal have the worst correlations, because these two indicators deny each other: WIF proposed by H&Y decreases by increasing of cited journal IF in previous year, but WIF proposed by G. Buela-Casal increases in the same case. The highest correlations are observed between average and median 5-year weighted impact factors.

**Table 5.** Pearson correlation coefficients between each pair of journal evaluation indicators.

| Indicators | JCR IF | Proposed WIF | WIF by H &Y | WIF by Buela-Casal | Average WIF | Median WIF | Average 5WIF | Median 5WIF |
|---|---|---|---|---|---|---|---|---|
| JCR IF | 1 | | | | | | | |
| Proposed | 0.7398 | 1 | | | | | | |
| WIF by H | 0.6120 | 0.7384 | 1 | | | | | |
| WIF by G. | 0.6105 | 0.6226 | 0.2797 | 1 | | | | |
| Average | 0.6256 | 0.8060 | 0.7414 | 0.6210 | 1 | | | |

| Median WIF | 0.6361 | 0.7233 | 0.6045 | 0.6421 | 0.8887 | I      |        |   |
|------------|--------|--------|--------|--------|--------|--------|--------|---|
| Average    | 0.7218 | 0.8135 | 0.8075 | 0.5774 | 0.9428 | 0.8647 | I      |   |
| Median     | 0.6932 | 0.7699 | 0.7473 | 0.5790 | 0.9489 | 0.8948 | 0.9474 | I |

## 5. Conclusion

As the special case from equation of proposed weighted impact factor we get current equation of impact factor calculated by ISI for JCR, but considering prestige and influence of cited journals gives more accurate indicator for journal ranking. Suggested method for IF can also apply to researcher, organization or country evaluation. Obviously, the paper or researcher cited by e.g. L. Zadeh is more important than object cited by unknown researcher. Therefore, regarding indexes of cited objects for research evaluation contributes more effective results.

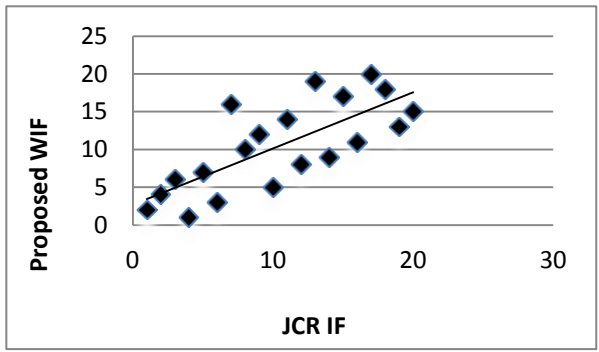

**Figure 1.** JCR IF versus proposed WIF

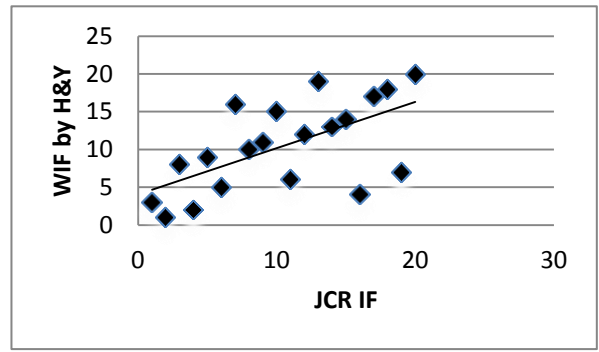

**Figure 2.** JCR IF versus WIF by H&Y

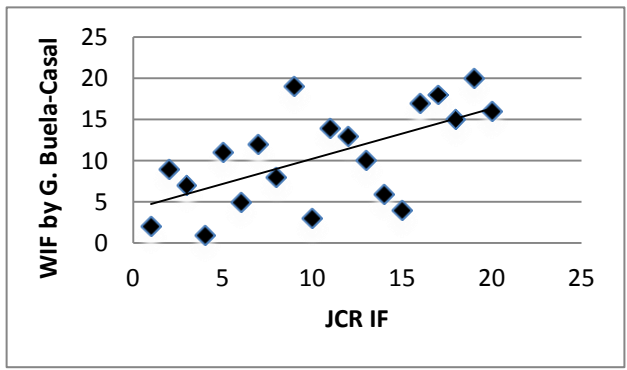

**Figure 3.** JCR IF versus WIF by G. Buela-Casal

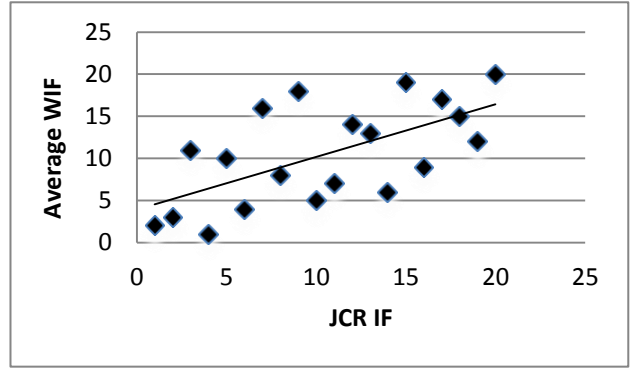

**Figure 4.** JCR IF versus Average WIF

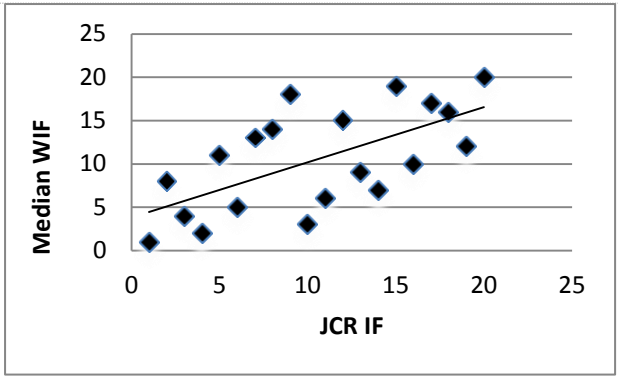

**Figure 5.** JCR IF versus Median WIF

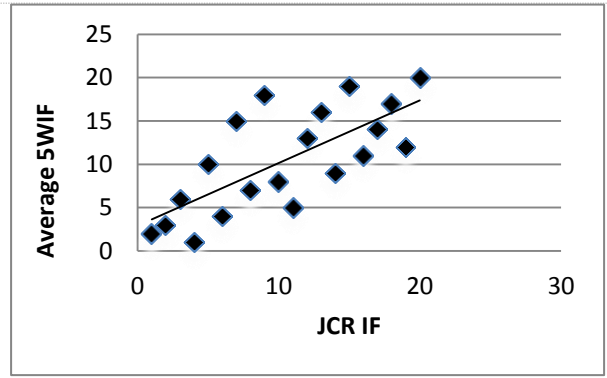

**Figure 6.** JCR IF versus Average 5WIF

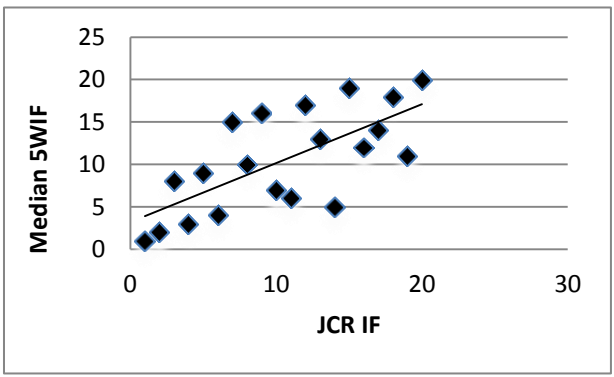

**Figure 7.** JCR IF versus Median 5WIF

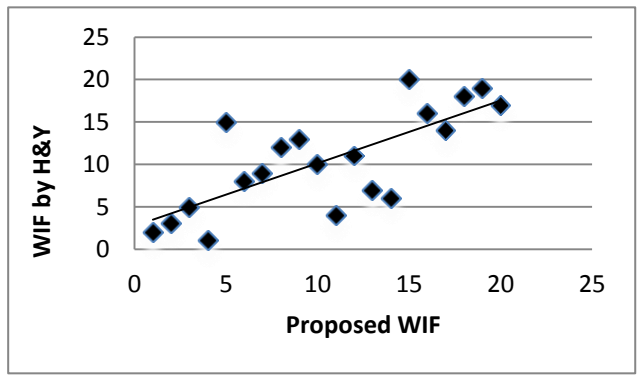

**Figure 8.** Proposed WIF versus WIF by H&Y

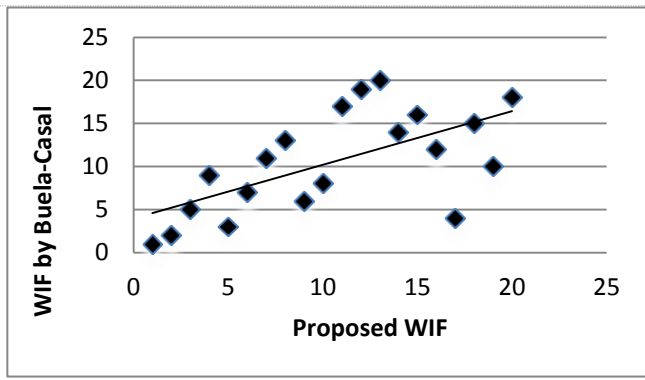

**Figure 9.** Proposed WIF versus WIF by Buela–Casal

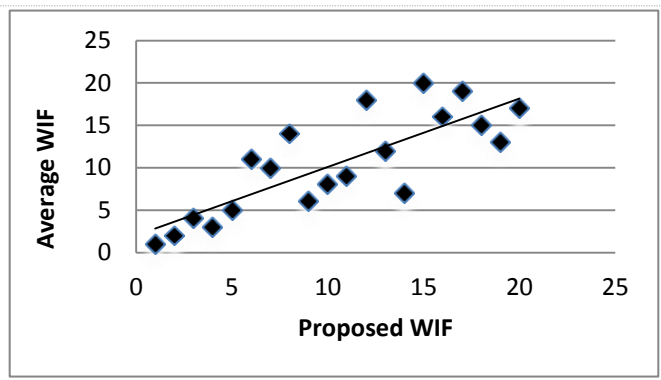

**Figure 10.** Proposed WIF versus Average WIF

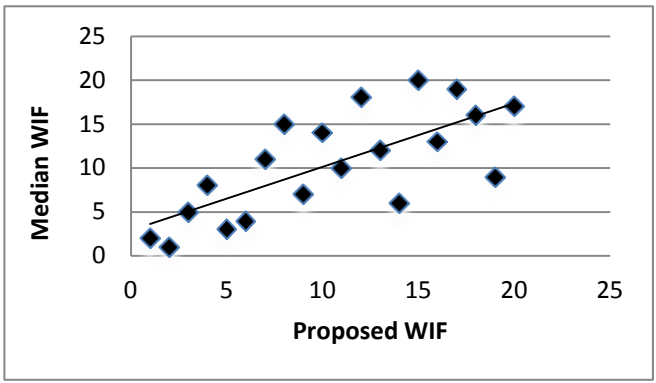

**Figure 11.** Proposed WIF versus Median WIF

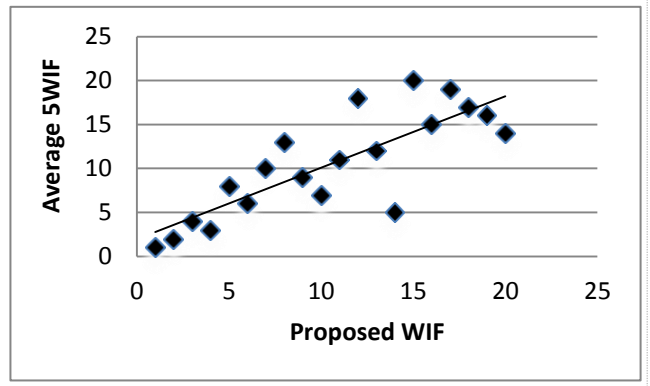

**Figure 12.** Proposed WIF versus Average 5WIF

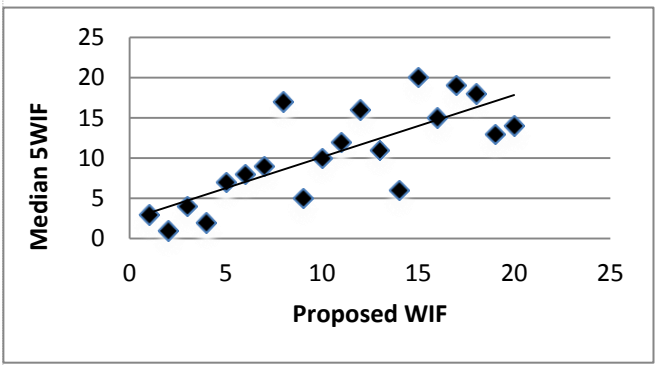

**Figure 13.** Proposed WIF versus Median 5WIF

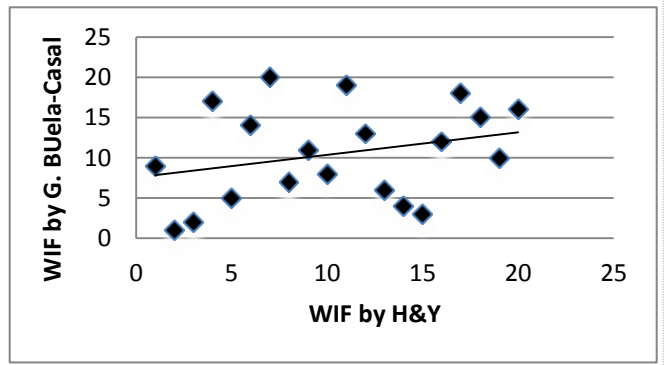

**Figure 14.** WIF by H&Y versus WIF by G. Buela-Casal

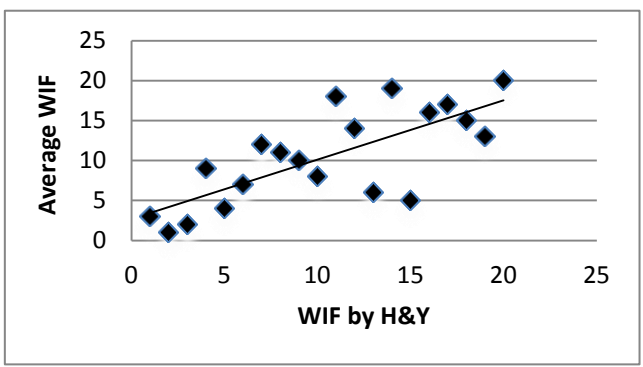

**Figure 15.** WIF by H&Y versus Average WIF

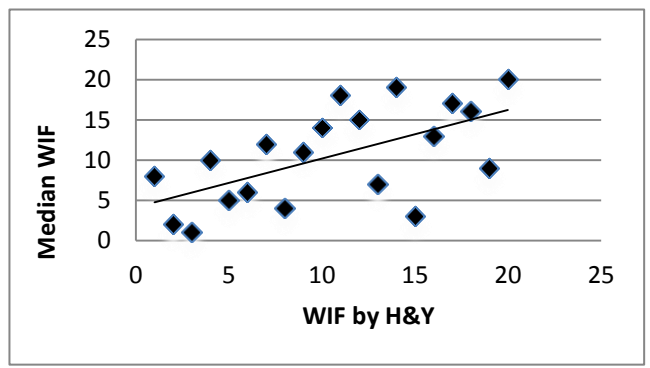

**Figure 16.** WIF by H&Y versus Median WIF

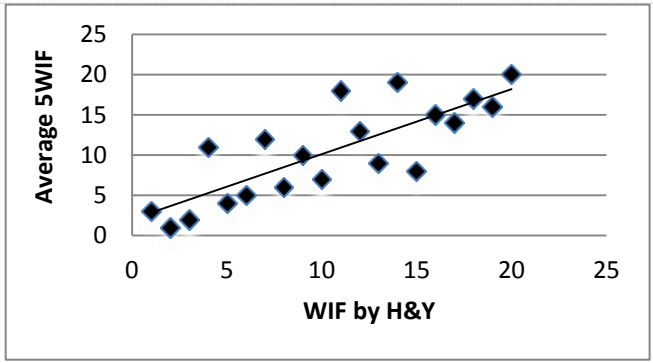

**Figure 17.** WIF by H&Y versus Average 5WIF

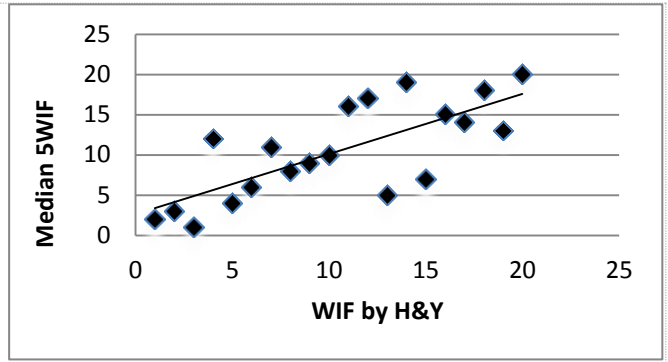

**Figure 18.** WIF by H&Y versus Median 5WIF

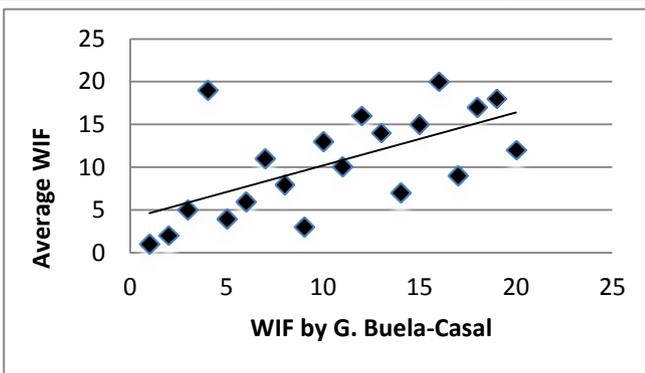

**Figure 19.** WIF by G. Buela-Casal versus Average WIF

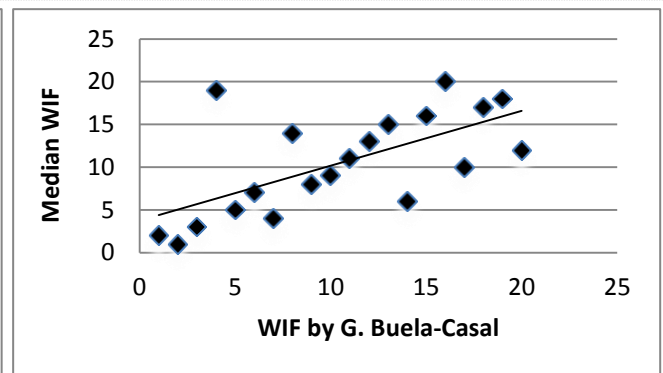

**Figure 20.** WIF by G. Buela-Casal versus Median WIF

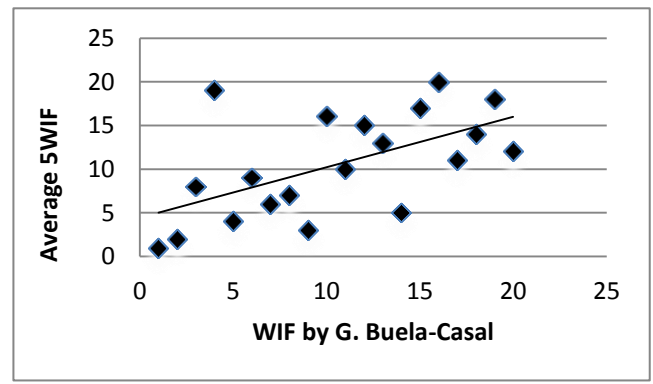

**Figure 21.** WIF by G. Buela-Casal versus Average 5WIF

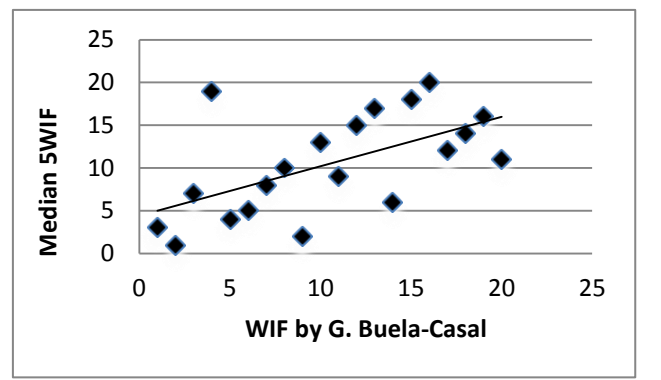

**Figure 22.** WIF by G. Buela-Casal versus Median 5WIF

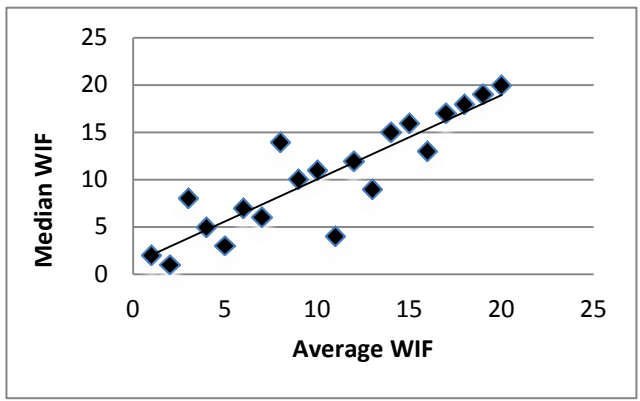

**Figure 23.** Average WIF versus Median WIF

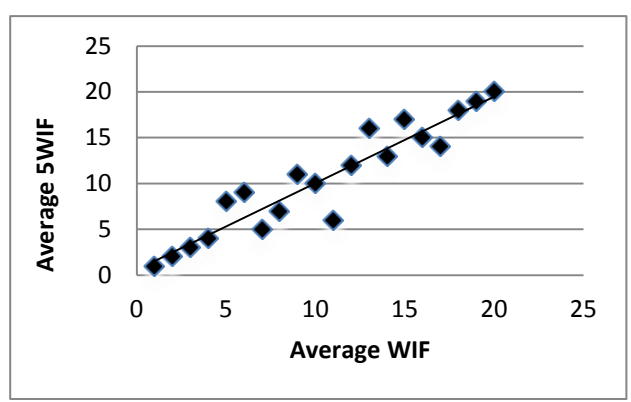

**Figure 24.** Average WIF versus Average 5WIF

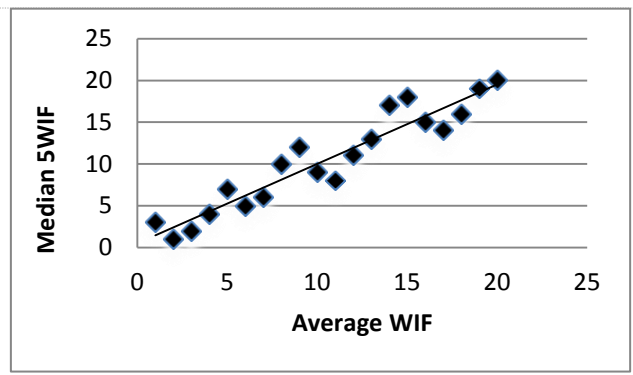

**Figure 25.** Average WIF versus Median 5WIF

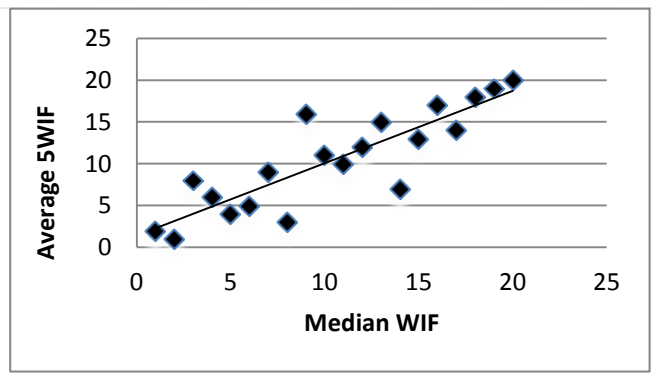

**Figure 26.** Median WIF versus Average 5WIF

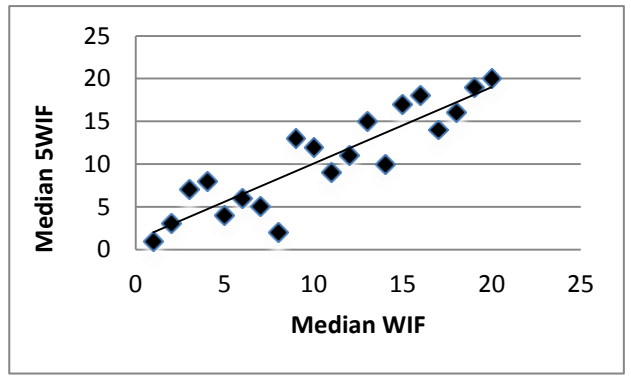

**Figure 27.** Median WIF versus Median 5WIF

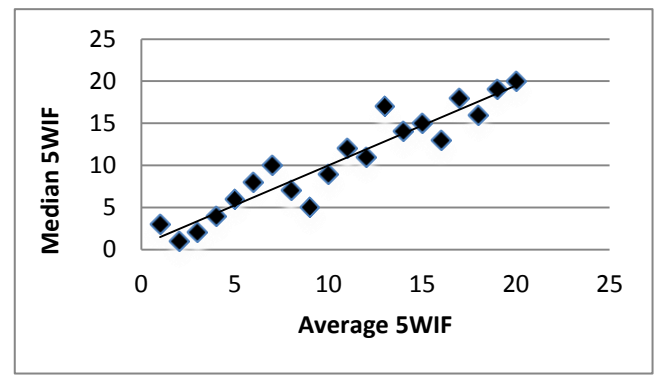

**Figure 28.** Average 5WIF versus Median 5WIF